# Site model of allometric scaling and fractal distribution networks of organs

Walton R. Gutierrez*
*Touro College, 27West 23rd Street, New York, NY 10010*
\*Electronic address: waltong@touro.edu

At the basic geometric level, the distribution networks carrying vital materials in living organisms, and the units such as nephrons and alveoli, form a scaling structure named here the site model. This unified view of the allometry of the kidney and lung of mammals is in good agreement with existing data, and in some instances, improves the predictions of the previous fractal model of the lung. Allometric scaling of surfaces and volumes of relevant organ parts are derived from a few simple propositions about the collective characteristics of the sites. New hypotheses and predictions are formulated, some verified by the data, while others remain to be tested, such as the scaling of the number of capillaries and the site model description of other organs.

## I. INTRODUCTION

**Important Note** (March, 2004): This article has been circulated in various forms since 2001. The present version was submitted to Physical Review E in October of 2003. After two rounds of reviews, a referee pointed out that aspects of the site model had already been discovered and proposed in a previous article by **John Prothero** ("Scaling of Organ Subunits in Adult Mammals and Birds: a Model" Comp. Biochem. Physiol. 113A.97-106.1996). Although modifications to this paper would be required, it still contains enough new elements of interest, such as **(P3)**, **(P4)** and **V,** to merit circulation. The q-bio section of www.arxiv.org may provide such outlet.

This article's starting point is the influential fractal model introduced by West *et al*. [1-3] of the internal circulation networks of organisms. Their model links geometric variables associated with a circulatory network to the "3/4-law" of basal metabolic rate. This aspect has received significant attention, but the issue of metabolic rate in animals entails various difficulties and controversies [1-11], perhaps the most serious of which is the fact that the "3/4-law" is not supported by the allometric data of reptiles, birds and large subsets of mammals [5,6,7]. On the other hand, many anatomic geometric variables of distribution networks follow allometric scaling relations, which as a whole, are certainly more important than a single metabolic rate rule. As the geometric variables are less prone to definition ambiguities than metabolic rate, their further study can lead to progress in this field. Towards this goal, the site model overcomes several of the difficulties of the current fractal model [1], as follows:
- It provides a framework for analysis of a large set of allometric variables, independent of the controversy surrounding metabolic rate laws.
- The allometric exponents of the geometric variables of distribution networks are logically derived from a small set of simple geometric principles.
- The distribution network can be quite general, and does not have to follow any special self-similarity rule (special fractals). Distribution network optimization analysis can still be applied in suitable cases [4,5,12].

- No special set of exponents plays a central role. A shortcoming of the fractal model [1] and many other general allometric models is the reliance on a particular set of exponents (y in $M_B^y$; $M_B$ = body mass) such as multiples of 1/4 or 1/3.
- It provides the first model of kidney allometry consistent with centralized distribution networks and in direct analogy to the lung. One of the attractive features of the site model is the possibility that other organs (liver, muscles, pancreas) may also be described by the same geometric principles of the site model.
- In the case of the lungs, it improves the overall fit of the modeled variables to the existing allometric data.

Whereas the model of West *et al.* [1] uses the blood capillary and its average surrounding volume as its basic unit, here the basic unit is the site represented by the nephron of the kidney and the alveolus of the lung. In the site model the blood capillary is one of the several components of a site, as in the nephron. In this way, the capillaries can play different geometric roles within each type of site associated with different organs.

In order to derive many specific geometric scaling laws of the organs the site model develops a simple geometric structure where most of the organ volume is divided into tiny sites, which are serviced by hierarchical distribution networks supplying air and blood. The model is summarized by four main propositions which are coupled with some known principles, such as mass conservation of the moving fluids in the networks. The resulting correlation to the allometric data of the kidney and lung of mammals is shown in Tables I-III. For example, the West *et al.* model predicts that the scaling of the total blood capillary volume in the lung is $M_B^{0.75}$, while the site model predicts it as $M_B^{0.96}$ and the data show $M_B^{1.0}$ (Table III).

## II. SITES, ALLOMETRY AND MASS CONSERVATION

This article makes extensive use of the scaling law between two variables X and Y, which, in order to avoid ambiguities, will be written as, $Y/Y' = (X/X')^d$, where the primed variables refer to another value of the variables without the prime. Taking $X' = 1$(unit), the scaling law becomes $Y = Y'X^d$ or $Y \propto X^d$, which is another equivalent form.

An organism or a particular organ is represented by a volume V. This is supplied by an internal distribution network of volume $V_{net}$ throughout the volume V. A large portion of the volume V, named $V_{TS}$, ($V > V_{TS}$), is subdivided by a large number $N_s$ of very small sites of volume $V_{1S}$ (see FIG.1), that is,

$$V_{TS}/N_s = V_{1S}. \qquad (1)$$

In the kidney a site volume $V_{1S}$ represents a nephron volume and in the lung the alveolus. This article focuses mainly on the kidney and lung of mammals, for which there is a more detailed allometric data for their parts.

West *et al.* [1] assume that the last relevant tiny units of the network are the blood capillaries, and the number of capillaries, $N_{cs}$ is found to scale with the volume V as,

$$N_{cs}/N_{cs}' = (V/V')^a \qquad (2)$$

with $a = 3/4$. Clearly, $N_{cs} > N_s$, since there are many capillaries in one site. However, there is no experimental evidence that $N_{cs}$ is proportional to $V^{0.75}$ or to $M_B^{0.75}$. Further complicating the matter is the fact that many capillaries control the blood flow with sphincter muscles. Thus, contrary to the assumption made by West *et al.* [1], the effective number of capillaries in use in a particular low (basal) metabolic state should be much smaller than the number of all capillaries ($N_{cs}$). The measurements of capillary density in muscle tissue are not conclusive and seem to favor an exponent greater than 0.75, as suggested by Dodds *et al*. [5] based on data of Hoppeler *et al*. [13]. Table III shows data of lung capillaries where clearly a > 0.75. In Section 3 a new theoretical geometric argument is made to derive an estimate for a. Although West *et al*. [1] and Turcotte *et al*. [14] make use of Eq. (2) in their justification of the "3/4-law," this article does not rely on any connection of Eq. (2) to metabolic rate laws. Furthermore, the validity of Eq. (2) can be argued by purely geometric methods without its possible relations to metabolic rate laws as suggested by Turcotte *et al*. [14]. However, their result correlates to just one lung, and hence lacks allometric validation, which requires a larger data sample of many species.

In direct analogy to Eq. (2), the site number $N_s$ is assumed to scale as,

$$N_s/N_s' = (V/V')^b, \qquad (3)$$

where the exponent b is determined later. While the volume of kidney V and its mass M should be proportional, it is not proportional to the body mass $M_B$, that is,

$$V/V' = (M_B/M_B')^c \qquad (4)$$

where c = 0.85 for the kidneys and c = 1.06 for the lungs of mammals. Other organs and body parts also have values of c in the approximate range of 1±0.30. These c exponents can be extrapolated from the data available for organ masses, summarized in Calder's book ([15], Tables 3-4, 5-3). However the direct proportion of mass and volume of a particular organ does not generally apply to the organs with significant air cavities. An important example is the lung, where, as it was given above, the lung volume of mammals scales as $M_B^{1.06}$, but the lung mass scales as $M_B$. Most of the allometric data relate the masses of body parts to body mass. This model shows that the geometric variables are important, therefore a direct correlation of the data on volumes, areas, and lengths of organ parts to the organ volume should be a very helpful step that is yet to be taken for most organs and phylogenic groups.

In the kidney, under the site interpretation given above, there is direct experimental evidence supporting Eq.(3). From the data ([15], Table 5-10), we have $N_s/N_s' = (M_B/M_B')^{0.62} = (M/M')^{0.62/0.85} = (V/V')^{0.73}$. That is, b = 0.73 in Eq. (3) in the case of the kidney.

The other basic component of the site model is the principle of mass conservation of moving fluids, also known as the equation of continuity, which for a fluid of constant density is

$$A_o U_o = N_k A_k U_k = N_s A_s U_s \qquad (5)$$

where the subscript k refers to any stage of the distribution network that can be numerically labeled in sequential order as k = 0, 1, 2, ... s, and $N_o$ = 1 (see FIG. 1). In the case of main arterial circulation, $U_o$ is the average speed of the blood in the aorta of cross section area $A_o$. There are $N_k$ pipe branches with average cross section $A_k$ where fluid moves with average flow speed $U_k$. The arterial bifurcation of the lung can be extended up to the capillaries (k = cs > s), but in the kidney, the nephron places the glomerulus before the last capillaries. Therefore, the last arterial network branching (k = s) is at the nephron level, which is the site in the kidney. If $A_{cs}$ is the cross section area of the capillaries where blood moves with the average speed $U_{cs}$, then $A_{cs}$ and $U_{cs}$ are approximately invariant ($M_B^0$) in most mammals. This is generally reinforced by the fact that red blood cells are size invariant and fit the capillaries very snugly.

A similar application of Eq. (5) can be made to the kidneys. In this case, $A_o$ is the average cross section area of the renal arteries. In the kidney, if the sites are represented by the nephrons, then $A_s$ and $U_s$ refer to the arteriole feeding the glomerulus which is the first filtration unit in the nephron that contains many capillaries and other structures. In this arteriole, the site flow $A_s U_s$ is not invariant ($M_B^0$), but scaling in relation to organ mass. From data of Table II, $A_o U_o$ is proportional to $M_B^{0.77}$ and $N_s$ is proportional to $M_B^{0.62}$; therefore, by using Eq. (5) we find $A_s U_s$ is proportional to $M_B^{0.12}$, which is quite different from $M_B^0$. In this last regard, the site model again departs from the model of West *et al.* [1], in which the last stage flow through the capillaries is invariant.

Now let us look at the scaling of $A_o \propto V^{b'}$. By looking at the data of the aorta, renal arteries, and trachea in Table I we can see a fairly constant exponent. The result obtained in this way is $b' = 0.725$. This is basically the same exponent observed for $N_s$ in the kidneys (above Eq. (5)). So it is proposed that $b' = b = 0.725$, for the major circulatory organs, that is,

$$A_o/A_o' = N_s/N_s' = (V/V')^b . \qquad (6)$$

Notice that by combining Eq. (6) and Eq. (5) we deduce that

$$A_o/N_s = A_s U_s /U_o = A_s' U_s' /U_o', \qquad (7)$$

which means that $A_s U_s /U_o$ is an invariant of the site model that can be experimentally tested beyond the case of the kidney. To summarize the site model so far, the following main propositions are introduced:

* **(P1)** Organ volume (V) is formed by a large number of sites ($V_{TS}$), such as the nephrons, and by hierarchical distribution networks ($V_{net}$). That is, $V = V_{TS} + V_{neta} + V_{netb} + ....$
* **(P2)** A geometric variable (Y) of the sites or distribution networks is invariant (x = 0), or scaling in relation to organ volume (V) for specific allometric groups. That is, $Y/Y' = (V/V')^x$.
* **(P3)** The cross section area ($A_o$) of the initial largest branch supplying the organ volume (V) is proportional to the number of sites ($N_s$). That is, $N_s/N_s' = A_o/A_o' = (V/V')^b$.

One more proposition (P4) is shown in the Section III. In proposition (P1) an allowance is made for possible multiple distribution networks such as the arterial, the venous, and the bronchial networks in the lung. All the variables defined around Eq. (5), as well as $V_{TS}$, and $V_{net}$, are assumed to be invariant or scaling in relation to V as described in (P2). Proposition (P3) states the intuitive notion that the number of site units forming an organ is proportional to the cross section size of the main branch supplying air or blood.

Because the supporting data in Tables I-III were obtained from different authors who employed heterogeneous methods, it is possible that these Tables are applicable only to larger terrestrial mammals. This article does not review the data comprehensively for any of the scaling laws under consideration. The higher accuracy of the exponents implied by the Tables could be interpreted as a tentative prediction of the site model until experimental measurements are improved. The experimental values quoted in the Tables are sometimes those closest to the theoretical value, but always from sources used repeatedly by prominent authors in this field. Current experimental errors are very difficult to uniformly evaluate due to the lack of standardized experimental and statistical procedures. For example, many allometric scaling laws are not more accurately established than $M_B{}^{y\pm 0.02}$ and a few may have larger error margins [15].

**III. SCALING SITES AND INVARIANT CAPILLARIES**

In this section various geometric scaling laws associated with the sites are derived and then followed by a discussion of the scaling of the number of blood capillaries. The sites are almost uniformly distributed in the volume $V_{TS}$ and occupy a large part of the organ volume V (about 80% or more), therefore we can expect that

$$V_{TS}/V_{TS}' = V/V'. \qquad (8)$$

This allows the derivation of several scaling laws associated with the sites which are very closely correlated to the experimental data. Combining Eq. (1), (3) and (8) we find

$$(V_{1S}/V_{1S}') = (V/V')^{1-b}. \qquad (9)$$

For example, the surface of one site scales as $V^{2(1-b)/3}$ and the surface of all sites $A_{TS}$ scales as

$$A_{TS}/A_{TS}' = (N_s/N_s')(V/V')^{2(1-b)/3} = (V/V')^{(2+b)/3}. \qquad (10)$$

Likewise the length of a site $L_{1S}$ scales as

$$L_{1S}/L_{1S}' = (V/V')^{(1-b)/3}. \qquad (11)$$

The application of these resulting scaling laws to the kidney and lung can be seen in Tables II, III. The scaling laws in relation to organ volume (V) are changed to body mass $M_B$ using Eq. (4).

Inside a site $V_{1S}$ there are other structures identified by the volume $V_{xs} < V_{1S}$ as in the examples of the glomerulus volume $V_{gs}$ in the nephron and $V_{cs}$ the volume of a blood capillary. The same scaling laws applicable to a site volume or surface also apply to any internal structure of a site that scales in the same way as the site. For example, the glomerulus volume scales approximately in the same way as the nephron (Table II) and the alveolus volume ($V_{as}$) scales like a lung circulation site (Table III). These last few relations can be summarized by: $V_{gs}/V_{gs}' = V_{as}/V_{as}' = V_{1S}/V_{1S}'$. The site scaling of the alveolus, however, does not preclude the existence of a larger lung site, such as the acinus, for which there are no allometric data. In this case, the alveolus volume ($V_{as}$) would be a scaling part of the acinus site, as the glomerulus is of the nephron.

A capillary volume in the nephron or in the alveolus is presumably invariant. To address the question of the number of capillaries, the following argument is made: since the exchange of materials by the capillaries is mainly a surface phenomenon, and the alveolar surface is uniformly covered with capillaries of constant cross section, then the total capillary number and volume in the lung would scale as the total alveolar surface. This is confirmed by the data [16] (see Table III). This leads us to the last main proposition of the site model:

\* **(P4)** The cross section area and volume of a capillary are invariant and the number of capillaries ($N_{cs}$) is proportional to the total site surface ($A_{TS}$). That is,

$$N_{cs}/N_{cs}' = A_{TS}/A_{TS}'. \qquad (12)$$

This last relation, combined with Eq. (10) and b = 0.725, sets the scaling of the total capillary number, length, surface, and volume in an organ as $M_B^{0.91c}$ or $V^{0.91}$. Here, it should be noted that in average muscle tissue where c = 1, the scaling of $M_B^{0.91}$ turns out to be quite different from the $M_B^{0.75}$ proposed by West *et al*. [1]. The capillary scaling proposed in the site model is, of course, only an approximation and may need revision in a more specific way for each organ. There are also a number of complicating factors in different animals, such as local capillary density, capillary wall thickness and blood oxygen partial pressure, to name a few [15,17].

The long-standing question of the scaling of the volume or mass of the kidney as $M_B^{0.85}$ can be addressed in the site model in the following way: if the most dominant geometric factor in the filtration work of the kidney is a total working surface ($M_B^{c(2+b)/3}$) proportional to the renal blood flow ($M_B^{0.77}$, Table II), then we conclude that 0.77 = c(2+b)/3, implying that c = 0.85. In other words, the kidney volume cannot scale with an exponent any larger or smaller than $M_B^{0.85}$ because there would be a mismatch between the renal blood flow and the total kidney working surface. It is noteworthy that up to this point, once b is set to 0.725, there are no additional "modeling" parameters in the site model.

## IV. OTHER ORGANS AND OTHER PHYLOGENIC GROUPS

Though it is well known that organs such as the liver and pancreas of mammals have an anatomic site structure, defined as the smallest unit performing a specific function of that organ, there are no allometric data for these sites. Similarity to the kidney site structure perhaps may be found in the liver, where its mass scales as $M_B^{0.87 \text{ to } 0.89}$ ([15], Table 3-4).

From the point of view of distribution networks, another important set of organs is the muscles, which we discuss briefly. Directionality in muscle tissue poses a difficult problem when trying to establish a simple scaling law of cross section capillary density in particular muscles, based on the site model or any other model. The variability of capillary density in muscles [13] is a sign of the complexity introduced by the necessary direction and type of motion exerted by different muscles. The site model may accommodate, to some degree, this variability in the following way: it is likely that different types of muscle are more like different types of organs, perhaps with total volume and cross section scaling in different ways in relation to body mass ($M_B^c$ with different values of c, see Eq. (4)). The hypothesized muscle site, which may be a fascicle, would change according to each type of muscle, as would the capillary geometry of the sites. Preliminary data confirm the approximation made by the site model for average muscle tissue. This is the scaling of total capillary length that Weibel *et al*. [18] estimates from data of mammals, with an exponent between 0.86 and 0.91, and which the site model predicts as 0.91 (c = 1, for many muscles). Reliable experimental data could clarify the roles of all these factors.

There are many allometric trends common to mammals and birds [15]. It is likely that a very similar site model is applicable to birds, but again the allometric data on distribution networks in this case are minimal. We find some supporting evidence in the trachea of birds [15], where the cross section ($A_o$) scales as $M_B^{0.70}$, and lung volume (V) as $M_B^{0.97}$, therefore $A_o$ scales as $V^{0.70/0.97} = V^{0.72}$, confirming for this case the value b = 0.725 given in Table I.

There are some allometric data on lung morphology of reptiles, but the samples are too small and there is no reference to the quality of the correlations [19].

## V. NETWORK OPTIMIZATION

Explaining allometric exponents by optimizing functions of a distribution network remains an important idea, albeit quite difficult to implement [5]. Metabolic rate optimization, as proposed by West *et al*. [1] is important, but unlikely to be the only important factor in the organs. This is already made clear by the variety of allometric exponents found in the scaling of the masses of the various organs in relation to body mass (Eq. (4)). We can identify many quantities subject to optimization in different organs: circulation distances, flow rates, area of exchange surfaces, total fluid volume, number of sites, size of sites, etc. Competing design requirements for a distribution network may also be evaluated with the help of computer simulations [3,20].

It is relevant to the site model to note that if total fluid circulation impedance is minimized [5] in a pulsatile regime (with other technical assumptions [5]), then the number of capillaries $N_{cs}$, scales as $V_{net}^{6/7} = V_{net}^{0.86}$, which is close to the $V^{0.91}$ proposed by the site model (if $V_{net}$ is proportional to V).

Another item of interest is the relation of the distribution network to the volume and form of the organ. It is clear that a strictly constant-ratio self-similar fractal is only a preliminary approximation to the networks found in organs [12, 21]. A notorious deficiency of such fractal distribution network models is that they convey very little information regarding the volume form of the organ where it is embedded. A recent article of Gutierrez [12] shows that it is possible to generalize the fractal characteristics of the network to include in its construction the details of the volume form of the organ. This construction yields a method to determine the volume of the network that in the first approximation uses the circulation distance function. When this function is chosen to scale optimally as $V^{1/3}$, then the volume of the network ($V_{net}$) scales at $V^{1.06}$. Combining this last observation with the scaling derived by Dodds *et al*. [5] for $N_{cs}$ as $V_{net}^{0.86}$ we find that $N_{cs}$ scales as $V^{0.86 \times 1.06} = V^{0.91}$ which is what the site model proposes!

**Acknowledgments.** Marie-Claude Vuille's assistance, discussion and encouragement have been invaluable in the preparation of this article.

TABLE I

Determination of b from $A_o \propto V^b$. Data from Calder [15], Table 5-3, p.92; Table 5-6, p.110; Table 5-10, p.131; or as indicated.

| Cross Section<br>Area ($A_o$) of | $A_o \propto M_B^{cb}$<br>cb | $M_B^c \propto V$=organ vol.<br>c | b |
|---|---|---|---|
| Trachea | 0.78 | 1.06 (lung) | 0.78/1.06=0.736 |
| Aorta | 0.72 | 1.0 (body) | 0.72/1.0 =0.72 |
| Renal artery | 0.61* | 0.85 (kidney) | 0.61/0.85=0.718 |

* Average of both [22].  <u>b = average b = 0.725</u>

TABLE II

Kidney geometric variables of mammals. Observed exp. is from Calder [15], Tables: 5-10, 5-11. No previous fractal model for kidney.  ND = No data, b = 0.725.

| Variable description | Obs. exp. | Site exp. | Model exp. | $M_B^{exp.} \propto$ Variable |
|---|---|---|---|---|
| Kidney Volume or mass | 0.85 | 0.85 | $=c=c_{kd}$ | V |
| Renal artery cross section | 0.61 | 0.62 | $c_{kd}b$ | $A_o$ |
| Number of nephrons (sites) | 0.62 | 0.62 | $c_{kd}b$ | $N_s$ |
| Number of glomeruli | 0.62 | 0.62 | $c_{kd}b$ | $N_s$ |
| Area per glomerulus | 0.18 | 0.16 | $2c_{kd}(1-b)/3$ | $V_{gs}^{2/3}$ |
| Capillaries per nephron | ND | 0.16 | $2c_{kd}(1-b)/3$ | $N_{cs}/N_s$ |
| Total glomerular surface | 0.73 | 0.77 | $c_{kd}(2+b)/3$ | $N_s V_{gs}^{2/3}$ |
| Total nephron volume | 0.85 | 0.85 | $c_{kd}$ | $V_{TS}$ |
| Total glomerular volume | 0.85 | 0.85 | $c_{kd}$ | $N_s V_{gs}$ |
| Number of blood capillaries | ND | 0.77 | $c_{kd}(2+b)/3$ | $N_{cs}$ |
| Renal arterial blood flow | 0.77 | 0.77 | $c_{kd}(2+b)/3$ | $A_o U_o$ |

TABLE III
Respiratory geometric variables of mammals. Previous model is from
West *et al*. [1]. Observed exp. is from Calder [15], see Tables
I and II, or as indicated. ND = No data.

| Variable description | Prev. model | Obs. exp. | Site Model exp. | exp. | $M_B^{exp.} \propto$ variable |
|---|---|---|---|---|---|
| Lung Volume | 1.0 | 1.06 | 1.06 | $=c$ | $V$ |
| Trachea cross section area | 0.75 | 0.78 | 0.77 | $cb$ | $A_o$ |
| Pulmonary artery cross section | 0.75 | ND | 0.77 | $cb$ | $A_o$ |
| Number of lung sites | 0.75 | ND | 0.77 | $cb$ | $N_s$ |
| Volume of alveolus or site | 0.25 | ND | 0.29 | $c(1-b)$ | $V_{as}$ |
| Radius of alveolus or site | 0.08 | 0.13 | 0.10 | $c(1-b)/3$ | $V_{as}^{1/3}$ |
| Surface of alveolus or site | 0.17 | ND | 0.19 | $2c(1-b)/3$ | $V_{as}^{2/3}$ |
| Alveolar area of lung | 0.92 | 0.95[&] | 0.96 | $c(2+b)/3$ | $N_s V_{as}^{2/3}$ |
| Capillary total area | 0.75 | 0.95[&] | 0.96 | " " | $N_{cs} A$ |
| Number of capillary in lung | 0.75 | ND | 0.96 | " " | $N_{cs}$ |
| Lung capillary total volume | 0.75 | 1.0[&] | 0.96 | " " | $N_{cs} V_c$ |

[&] Gehr *et al*. [16].

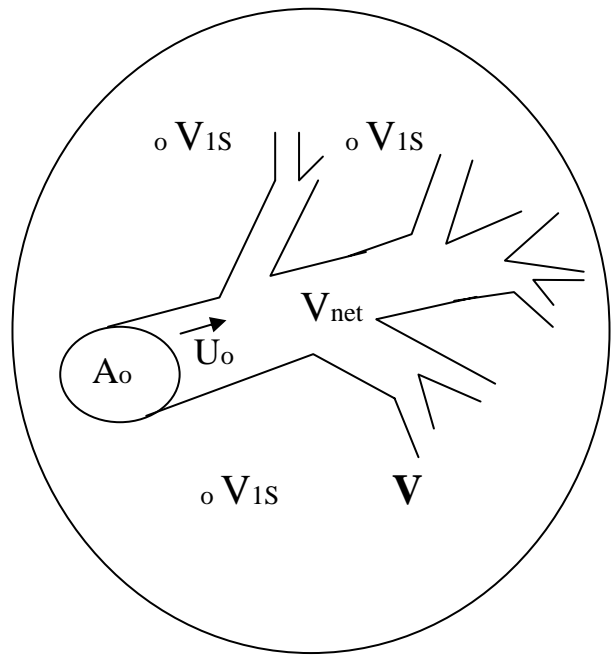

FIG. 1. Schematics of volume V showing some network variables of the site model.

Walton Gutierrez          SM56D7